\documentclass[twocolumn,trackchanges]{aastex701}
\usepackage{graphicx}

\usepackage{multirow}
\shorttitle{Robust TRGB magnitude}
\shortauthors{Chung et al.}
\graphicspath{{./}{figures/}}

\begin{document}

\title{How Robust is the Cosmic Distance with Tip of Red Giant Branch against \\Stellar Population Variations?}

\author[0000-0001-6812-4542]{Chul Chung}
\affil{Department of Astronomy \& Center for Galaxy Evolution Research, Yonsei University, Seoul 03722, Republic of Korea}
\email[show]{chulchung@yonsei.ac.kr}

\author[0000-0002-2210-1238]{Young-Wook Lee}
\affil{Department of Astronomy \& Center for Galaxy Evolution Research, Yonsei University, Seoul 03722, Republic of Korea}
\email{ywlee2@yonsei.ac.kr}

\author[0000-0002-1842-4325]{Suk-Jin Yoon}
\affil{Department of Astronomy \& Center for Galaxy Evolution Research, Yonsei University, Seoul 03722, Republic of Korea}
\email{sjyoon0691@yonsei.ac.kr}

\author[0000-0002-1907-0848]{Yong -Cheol Kim}
\affil{Department of Astronomy, Yonsei University, Seoul 03722, Republic of Korea}
\email{yckim@yonsei.ac.kr}

\author{Sang-Il Han}
\affil{Department of Science Education, Ewha Womans University, Seoul 03760, Republic of Korea}
\email{sangil.han@gmail.com}

\author[0000-0001-5966-5072]{Hyejeon Cho}
\affil{Department of Astronomy \& Center for Galaxy Evolution Research, Yonsei University, Seoul 03722, Republic of Korea}
\email{hyejeon@yonsei.ac.kr}

\author[0000-0001-7277-7175]{Dongwook Lim}
\affil{Department of Astronomy \& Center for Galaxy Evolution Research, Yonsei University, Seoul 03722, Republic of Korea}
\email{dwlim@yonsei.ac.kr}

\author[0000-0002-1031-0796]{Young-Lo Kim}
\affil{Department of Astronomy \& Center for Galaxy Evolution Research, Yonsei University, Seoul 03722, Republic of Korea}
\email{ylkim83@yonsei.ac.kr}

\author[0000-0002-1562-7557]{Sohee Jang}
\affil{Department of Astronomy \& Center for Galaxy Evolution Research, Yonsei University, Seoul 03722, Republic of Korea}
\email{sohee.jang@yonsei.ac.kr}

\author[0000-0003-4364-6744]{Seungsoo Hong}
\affil{Department of Physics and Astronomy, Seoul National University, Seoul 08826, Republic of Korea}
\email{sshong@snu.ac.kr}

\author[0009-0002-1351-1582]{Seunghyun Park}
\affil{Department of Astronomy \& Center for Galaxy Evolution Research, Yonsei University, Seoul 03722, Republic of Korea}
\email{daelikii@gmail.com}

\author[0009-0004-3117-1977]{Junhyuk Son}
\affil{Department of Astronomy \& Center for Galaxy Evolution Research, Yonsei University, Seoul 03722, Republic of Korea}
\email{sonjunhyuk@yonsei.ac.kr}

\author[0000-0003-2713-6744]{Myung Gyoon Lee}
\affil{Astronomy Program, Department of Physics and Astronomy, SNUARC, Seoul National University, 1 Gwanak-ro, Gwanak-gu, Seoul 08826, Republic of Korea}
\email{mglee@astro.snu.ac.kr}

\begin{abstract}
The tip of the red giant branch (TRGB) provides a key standard candle for extragalactic distance measurements and for refining the Hubble constant.
We test its robustness by quantifying how metallicity, $\alpha$-element enhancement, age, and initial helium abundance modulate the TRGB luminosity, using synthetic composite color--magnitude diagrams in the $I$ and $F814W$ bands.
We find that metallicity and $\alpha$-element enhancement are the primary drivers of TRGB variation, while age introduces only a modest effect and helium abundance is negligible.
At fixed age and helium content, increasing the mean metallicity by 0.5 dex or the $\alpha$-element enhancement by 0.3 dex produces the well-known systematic dimming of 0.046 and 0.050 mag, respectively, in $M_I^{\rm TRGB}$, and of 0.093 and 0.044 mag, respectively, in $M_{F814W}^{\rm TRGB}$. 
By comparison, changes in age of 3~Gyr and in initial helium abundance of 0.10 yield  minor luminosity shifts, with average changes of 0.031 and 0.009~mag, respectively, in $M_I^{\rm TRGB}$, and of 0.035 and 0.027 mag, respectively, in $M_{F814W}^{\rm TRGB}$, substantially smaller than those caused by variations in metallicity or $\alpha$-element enhancement. 
For mixed stellar populations under typical stellar-halo metallicity conditions, the net variation in $M_I^{\rm TRGB}$ arising from each combination of the $\alpha$-element enhancement, age, and initial helium abundance remains below 0.028~mag, well within reported systematic uncertainties.
Together, these results reaffirm the TRGB as a highly robust distance
indicator and support its continued use as an independent anchor for
precision cosmology in the era of the Hubble-tension debate.
\end{abstract}


\section{Introduction}
\label{s1}

The tip of the red giant branch (TRGB) magnitude is a well-known distance indicator, marking the peak luminosity achieved by low-mass stars ($M \leq 2M_\odot$) during their transition from hydrogen shell burning to helium core ignition in the red giant branch (RGB) phase \citep{1993ApJ...417..553L, 1997MNRAS.289..406S, 2008MmSAI..79..440B, 2012MNRAS.427..127B, 2017A&A...606A..33S}.
Observationally, the TRGB is identified as a sharp cutoff in the luminosity function of RGB stars, particularly prominent in the $I$-band and near-infrared \citep[e.g.,][]{1993ApJ...417..553L, 2006AJ....132.2729M, 2009ApJ...690..389M, 2017ApJ...835...28J, 2017ApJ...845..146H, 2019ApJ...880...63M}.
This well-defined feature makes the TRGB magnitude a robust standard candle for measuring extragalactic distances \citep[e.g.,][{and references therein}]{2017ApJ...836...74J, 2020ApJ...891...57F, 2021ApJ...906..125J, 2023AJ....166....2M}.

The TRGB magnitude is often regarded as nearly universal in the $I$-band and near-infrared, as supported by both observational and theoretical studies \citep[e.g.,][]{1993ApJ...417..553L, 2017ApJ...835...28J, 2020ApJ...891...57F}. 
In practice, the metallicity dependence of the TRGB has been extensively addressed through color-based calibrations \citep[e.g.,][]{2009ApJ...690..389M, 2008MmSAI..79..440B, 2017ApJ...835...28J}, and the effects of other stellar population parameters such as $\alpha$-element enhancement \citep[e.g.,][]{1997MNRAS.289..406S, 2002ApJS..143..499K, 2023AJ....166....2M} and age \citep{2008MmSAI..79..440B, 2012MNRAS.427..127B} have also been explored in previous studies. 
However, a comprehensive analysis that simultaneously examines these parameters within a self-consistent synthetic color–magnitude diagram (CMD) framework is still lacking. 
As a result, the combined influence of stellar population parameters, together with variations in the initial helium abundance, on the TRGB luminosity in synthetic CMDs has not yet been systematically quantified. This limitation is particularly relevant for TRGB magnitude studies, since the TRGB is commonly measured in galactic halos whose structure and composition reflect complex mixtures of stellar populations arising from hierarchical merging, in situ star formation, and dynamical interactions \citep[e.g.,][]{1985AJ.....90.2089B, 2003MNRAS.339..897T, 2005ApJ...621L..57L, 2008ApJ...689..936J, 2010MNRAS.406..744C}.
Stellar halos typically contain low-metallicity stars originating from accreted dwarf galaxies \citep[e.g.,][]{2006ApJ...642L.137B, 2018Natur.563...85H} and exhibit a range of $\alpha$-element enhancements, with radial trends indicating a more complex chemical enrichment history. 
In addition, the age structure of stellar halos, shaped by accretion and star formation history \citep[e.g.,][]{1978ApJ...225..357S, 1994ApJ...423..248L}, also significantly influence the stellar populations of halos. 
The presence of helium-rich stars in Milky Way globular clusters such as $\omega$~Cen and NGC~2808 \citep[e.g.,][]{2005ApJ...621L..57L, 2007ApJ...661L..53P, 2011ApJ...728..155D, 2011A&A...531A..35P} may further contribute to stellar population variations in the halo and outer fields by dispersing helium-rich stars into their environments.
These findings highlight how chemical enrichment and star formation histories imprint systematic variations onto stellar populations, which may in turn influence the TRGB magnitude and affect precision distance measurements.

In this paper, we investigate how metallicity, $\alpha$-element enhancement, age, and initial helium abundance impact the TRGB magnitude and assess the implications for distance determination and cosmology.
In Section~\ref{s2}, we compare isochrones with varying parameters and provide theoretical explanations for the resulting TRGB magnitudes.
Section~\ref{s3} presents synthetic CMDs and examines the corresponding TRGB magnitude in Johnson-Cousins $I$ and HST ACS/WFC $F814W$ bands.
In particular, we investigate how the mixing of different composite stellar populations affects variations in the TRGB magnitude.
In Section~\ref{s4}, we discuss how these population effects influence the inferred value of Hubble constant ($H_0$) and propose new considerations for future cosmological studies.

\section{The `Tip of RGB' magnitudes from Isochrones}
\label{s2}

\begin{figure}
\centering
\includegraphics[angle=-90,scale=0.46]{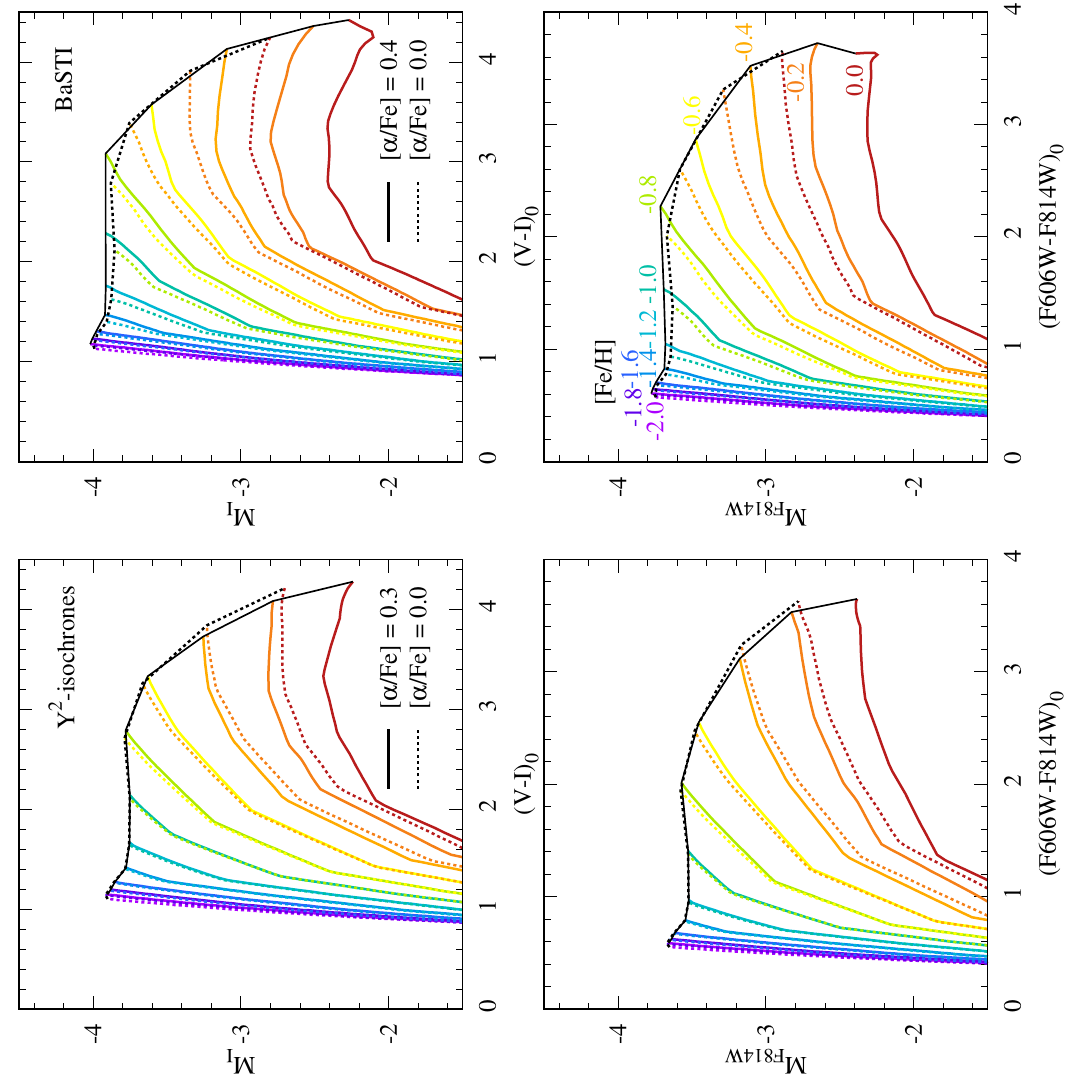}
\caption{
The effect of ${\rm [Fe/H]}$ and $[\alpha/{\rm Fe}]$ on the TRGB magnitudes in the Johnson-Cousins $I$ (upper) and HST ACS/WFC $F814W$ (lower) bands. 
Two sets of theoretical 12~Gyr isochrones are compared, with the $Y^2$-isochrones (left) and the BaSTI isochrones (right). 
Different colors represent varying metallicities, while solid and dashed lines correspond to $[\alpha/{\rm Fe}] = 0.3$ and 0.0, respectively.
For the BaSTI isochrones, the $\alpha$-element enhanced models correspond to $[\alpha/{\rm Fe}] = 0.4$.
The horizontal black lines connect the TRGB magnitudes at different metallicities to the corresponding ${\rm [\alpha/Fe]}$ line types in each CMD.
The predicted TRGB magnitudes exhibit some variation with $[\alpha/{\rm Fe}]$ within each set of isochrones.
}
\label{f1}
\end{figure}

\begin{figure}
\centering
\includegraphics[angle=-90,scale=0.46]{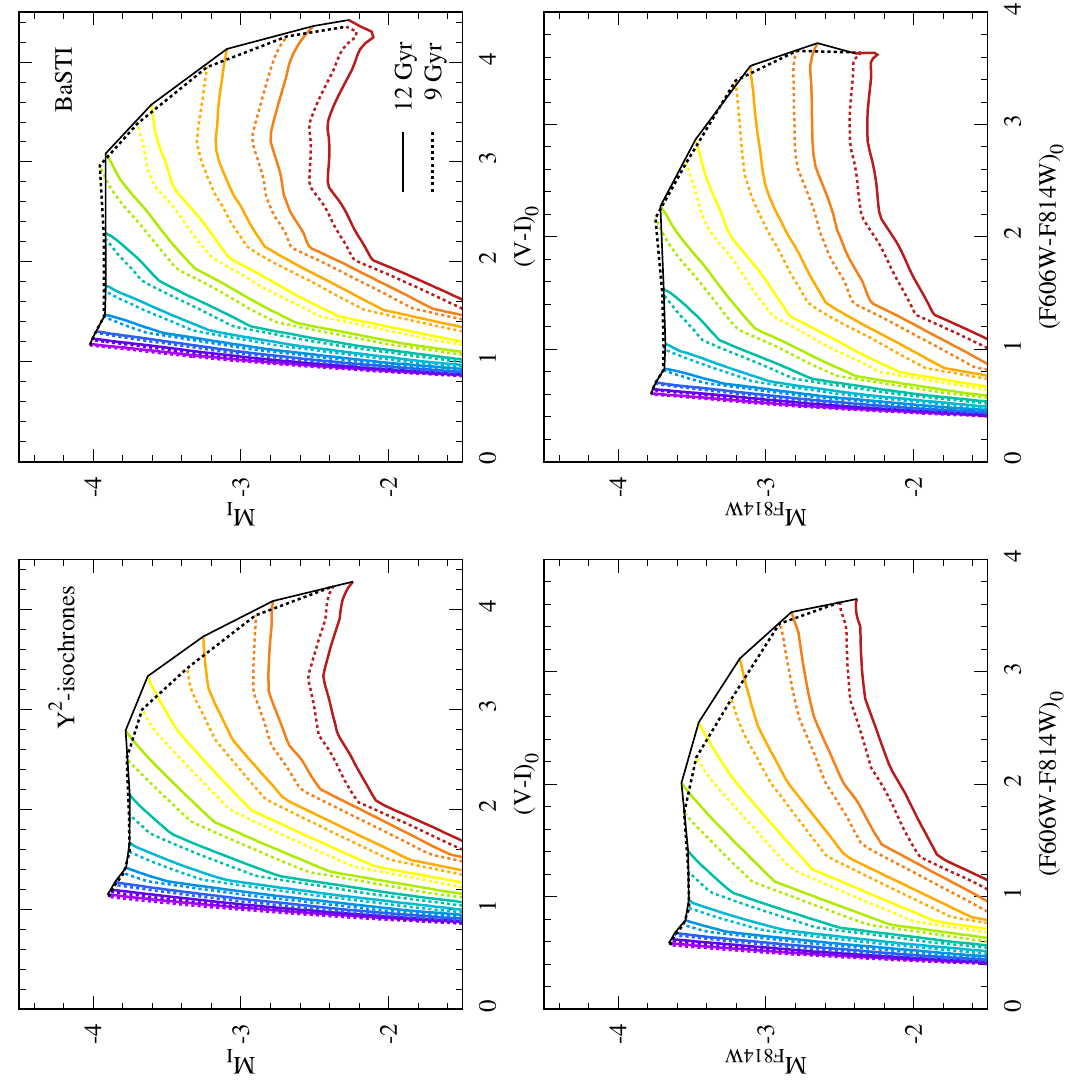}
\caption{
Same as Figure~\ref{f1}, but showing the effect of age differences at fixed $\alpha$-enhancement, with $[\alpha/{\rm Fe}]=0.3$ and 0.4 for the $Y^2$ and BaSTI isochrones, respectively.
Solid and dashed lines represent isochrones for 12 and 9~Gyr, respectively. 
The younger (9 Gyr) isochrones exhibit bluer colors than their older (12 Gyr) counterparts, resulting in slightly brighter TRGB magnitudes for metal-poor stars.
}
\label{f2}
\end{figure}

\begin{figure}
\centering
\includegraphics[angle=-90,scale=0.46]{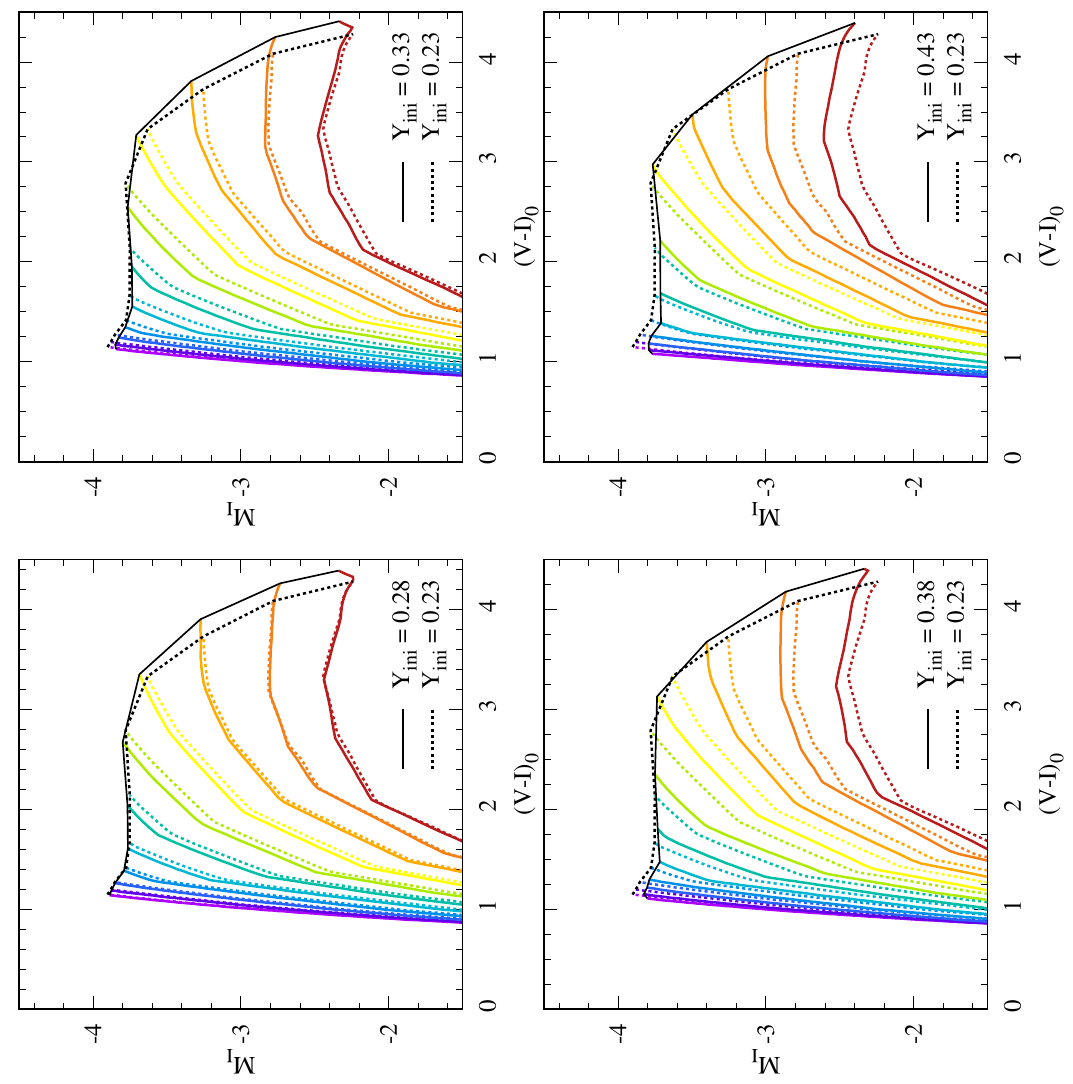}
\caption{
The effect of helium abundance on the TRGB magnitude of $Y^2$-isochrones in $(V-I)_0$ versus $M_I$.
The rainbow colors indicate the same metallicities as in Figure~\ref{f1}, while the age and ${\rm [\alpha/Fe]}$ are fixed at 12~Gyr and 0.3.
Dashed lines represent normal-helium isochrones with $Y_{\rm ini} = 0.23$, while solid lines in each panel correspond to helium-enhanced isochrones with $Y_{\rm ini} = 0.28$, 0.33, 0.38, and 0.43, as indicated in the bottom right. 
The TRGB magnitudes of helium-enhanced isochrones with ${\rm [Fe/H]} \leq -1.6$ are slightly fainter than those of the normal-helium isochrones at the same metallicity.
}
\label{f3}
\end{figure}

\begin{figure}
\centering
\includegraphics[angle=-90,scale=0.46]{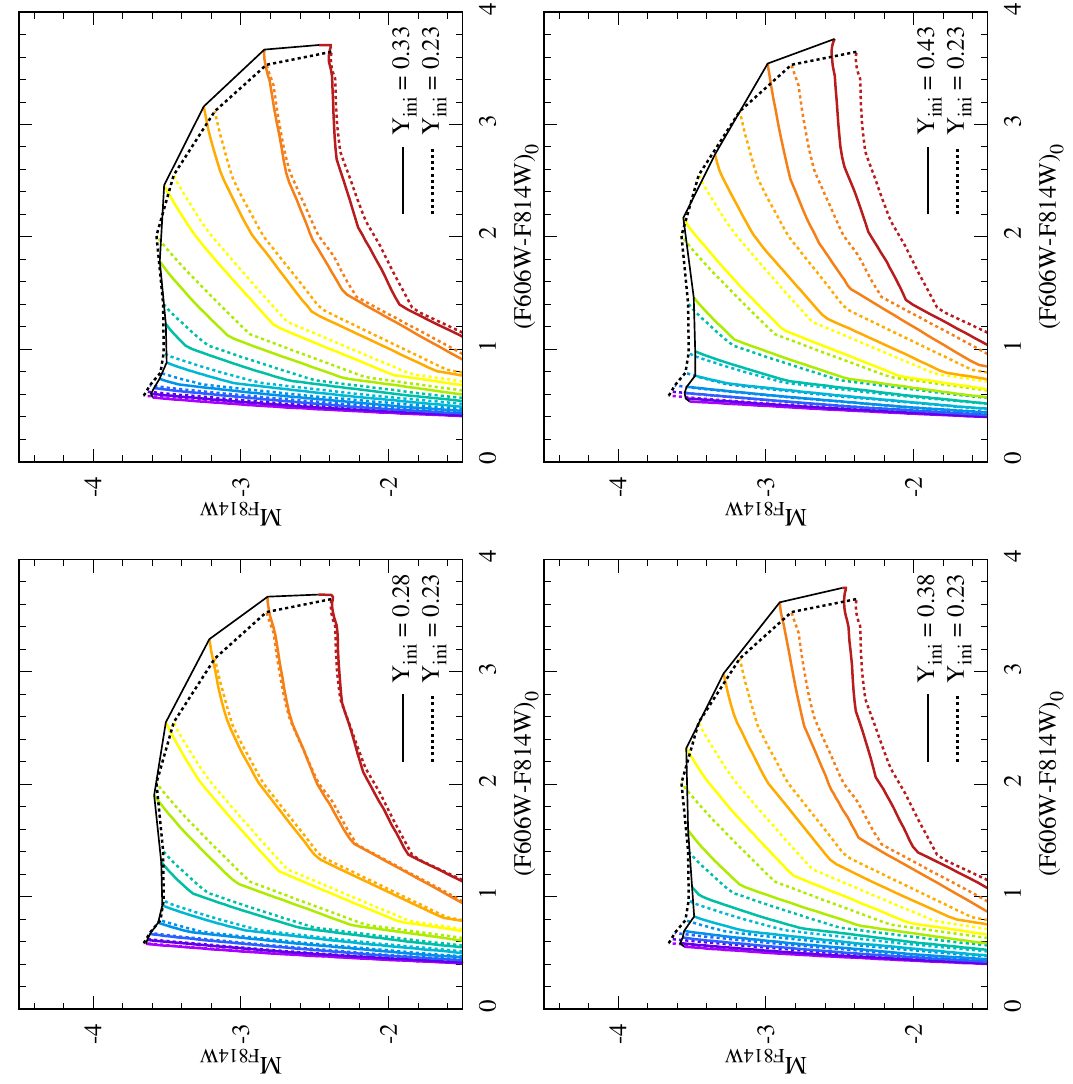}
\caption{
Same as Figure~\ref{f3}, but for $({{F606W} - F814W})_0$ versus $M_{F814W}$. 
Similar to the behavior in the $I$ band, the TRGB magnitudes in $M_{F814W}$ for ${\rm [Fe/H]} \leq -1.6$ become slightly fainter with increasing initial helium abundance in the helium-enhanced isochrones compared to those in the normal-helium ones.
}
\label{f4}
\end{figure}

\begin{figure*}
\centering
\includegraphics[angle=-90,scale=0.7]{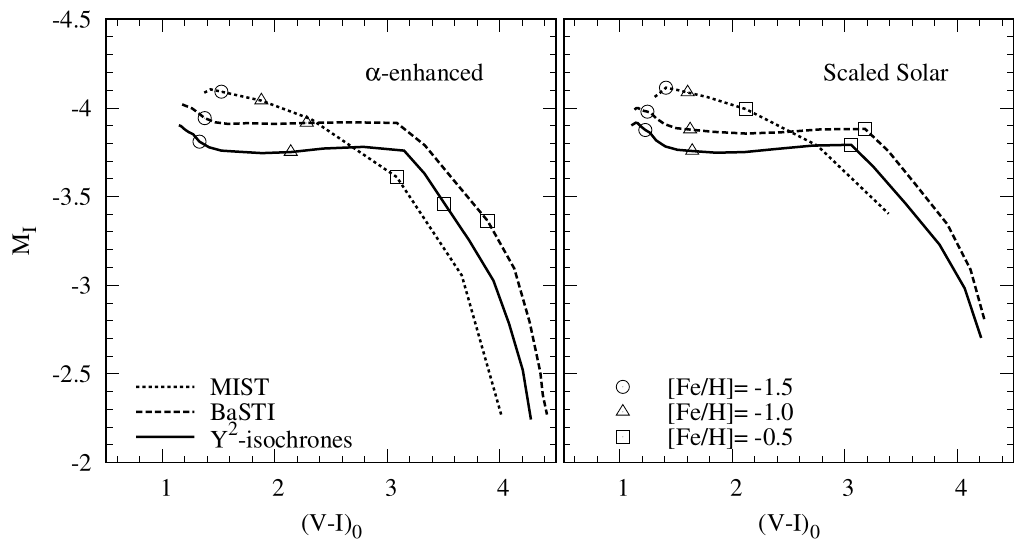}
\caption{
Comparison of the TRGB magnitudes from the Y$^2$, BaSTI, and MIST isochrones in the $(V-I)_0$ versus $M_I$ plane. 
The left and right panels show the $\alpha$-enhanced and scaled-solar cases, respectively. 
The adopted $\alpha$-enhancements are $[\alpha/{\rm Fe}]=0.3$, 0.4, and 0.4 for the Y$^2$, BaSTI, and MIST isochrones, respectively.
The Y$^2$ and BaSTI isochrones are plotted for an age of 12 Gyr, whereas the MIST isochrones are shown for 12.5 Gyr.
The TRGB magnitudes are plotted over ${\rm [Fe/H]}=-2.0$ to $0.0$, with the points at ${\rm [Fe/H]}=-1.5$, $-1.0$, and $-0.5$ marked by open circles, triangles, and squares, respectively. 
Despite slight differences in the MIST isochrones, the overall trend of fainter TRGB luminosity with increasing metallicity is common to all model sets.
}
\label{f5}
\end{figure*}

\begin{figure*}
\centering
\includegraphics[angle=-90,scale=0.66]{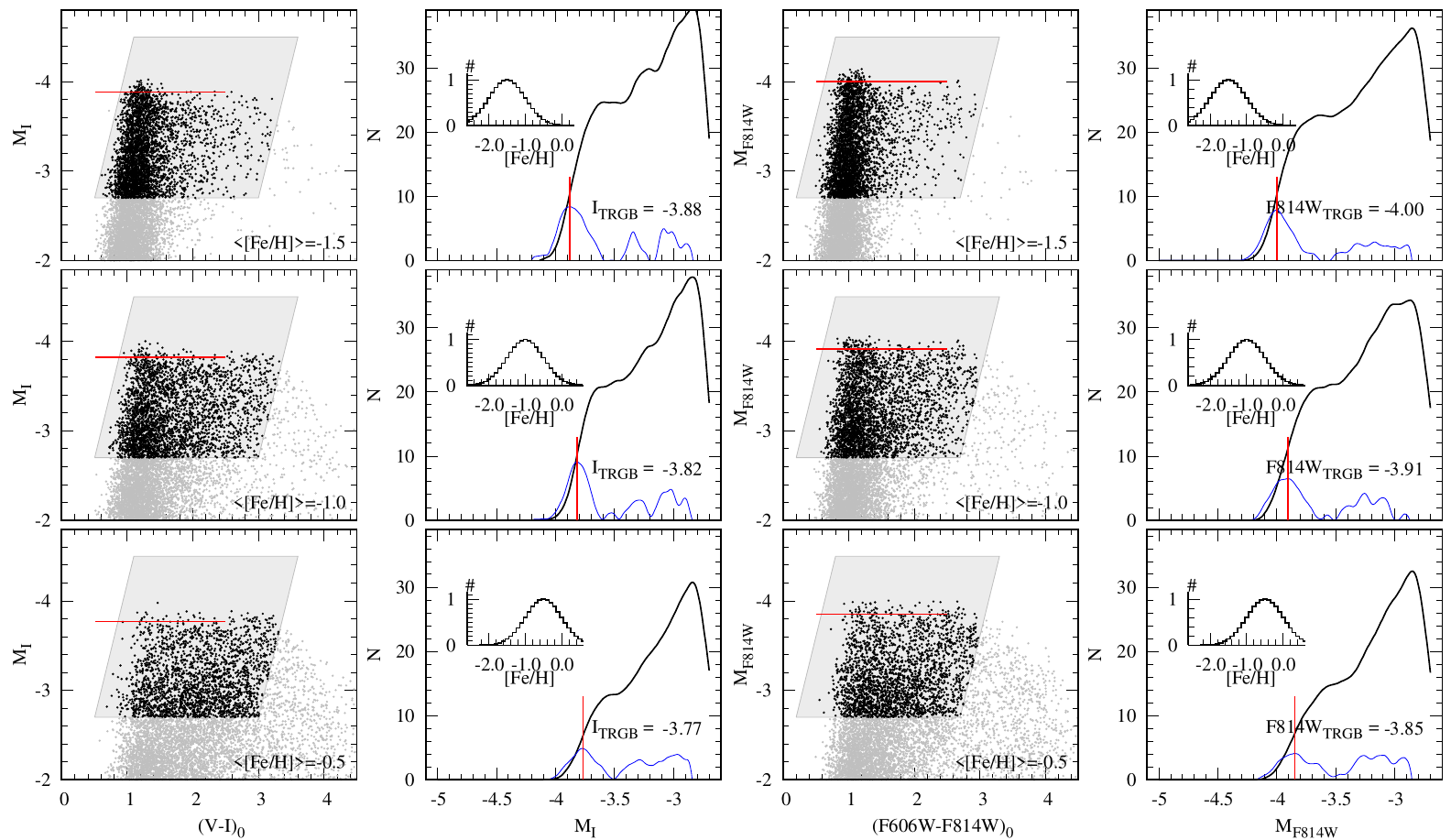}
\caption{
{Examples of synthetic composite CMDs and corresponding luminosity functions at different mean metallicities.}
$M_{I}^{\rm TRGB}$ and $M_{F814W}^{\rm TRGB}$ magnitudes derived from synthetic $(V-I)_0$ versus $M_I$ and $({{F606W} - F814W})_0$ versus $M_{F814W}$ CMDs with varying metallicity distributions are presented.
The other stellar parameters, ${\rm [\alpha/Fe]}$, age, and $Y_{\rm ini}$, are fixed at 0.3, 12~Gyr, and 0.23, respectively.
First and third columns: Synthetic composite CMDs based on $Y^2$-isochrones under different mean metallicity distributions ($\left< {\rm [Fe/H]} \right>$) with $\sigma_{\rm \left< [Fe/H] \right>} = 0.5$ assumptions.
Gray-shaded parallelograms mark the TRGB selection regions.
Red lines indicate the TRGB magnitudes detected by the Sobel-edge detection in the corresponding even column panels.
Second and fourth columns: Luminosity functions within the selection box, with blue lines showing Sobel-edge detections and red lines indicating $M_{I}^{\rm TRGB}$ and $M_{F814W}^{\rm TRGB}$. 
Insets display the input Gaussian metallicity distributions.
The CMDs and Sobel-edge detections shown are one example out of 100 simulations. 
Both $M_{I}^{\rm TRGB}$ and $M_{F814W}^{\rm TRGB}$ magnitudes become fainter with increasing mean metallicity of the MDF.
}
\label{f6}
\end{figure*}

\begin{figure*}
\centering
\includegraphics[angle=-90,scale=0.66]{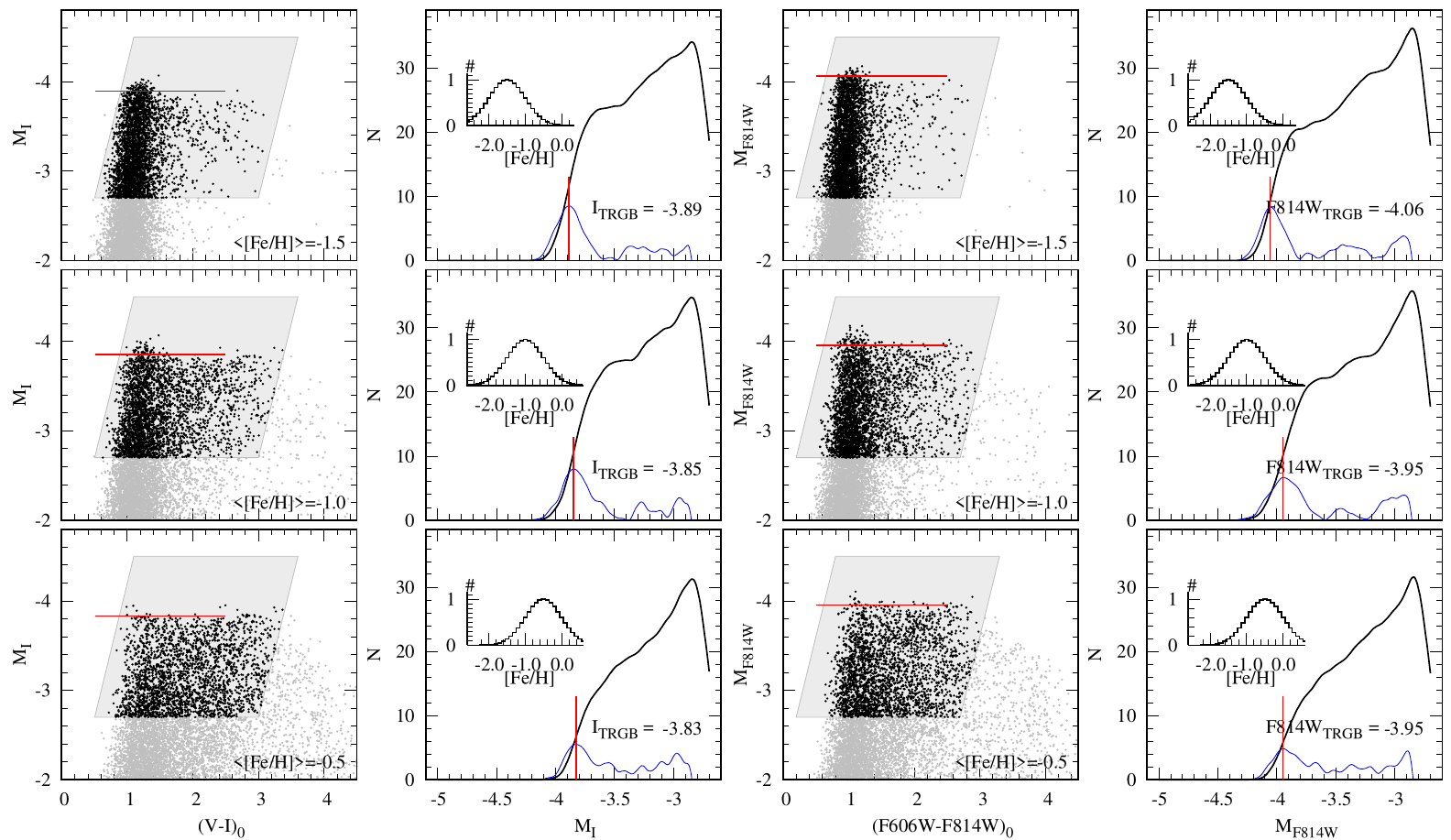}
\caption{
Same as Figure~\ref{f6}, but for ${\rm [\alpha /Fe]}=0.0$. 
The TRGB magnitudes from the scaled-solar model predict slightly brighter $M_I$ and $M_{F814W}$ values than those from the $\alpha$-enhanced models.
}
\label{f7}
\end{figure*}

The intrinsic luminosity of the TRGB is closely linked to the helium core mass, with a reported scaling relation of $\partial \log (L_{\rm TRGB}/L_{\odot}) / \partial \log (M_c^{\rm He} / M_{\odot}) \sim 6$, where $L_{\rm TRGB}$ is the bolometric luminosity at the TRGB and $M_c^{\rm He}$ is the helium core mass at the onset of helium ignition \citep[][see their Figure~1]{2017A&A...606A..33S}. 
Therefore, any change in TRGB luminosity is primarily driven by variations in the helium core mass. 
However, from an observational perspective, the metallicity-dependent opacity in stellar atmospheres also plays a significant role by affecting flux absorption and altering the observed TRGB magnitude \citep[e.g.,][]{2017A&A...606A..33S}. 
It is thus essential to account for both the core mass dependence and atmospheric opacity effects when interpreting TRGB luminosity variations.

Figure~\ref{f1} presents the TRGB magnitudes predicted by the theoretical models of the $Y^2$-isochrone \citep{2002ApJS..143..499K} and BaSTI isochrones \citep{2006ApJ...642..797P} in the Johnson-Cousins $I$ and HST ACS/WFC $F814W$ passbands.
These passbands are the most commonly used and well-calibrated for TRGB magnitude in distance measurement studies.
For the ACS/WFC filters, we adopt ${F606W}$ as the HST counterpart to the V band.
The figure includes $\alpha$-enhanced models with ${\rm [\alpha/Fe]} = 0.0$ and $0.3$ for the $Y^2$-isochrones, and ${\rm [\alpha/Fe]} = 0.0$ and $0.4$ for the BaSTI isochrones.
Both sets of isochrones exhibit a clear cutoff in the $I$ and $F814W$ band magnitudes around $-4.0$~mag, primarily formed by metal-poor populations with ${\rm [Fe/H]} < -0.8$.
The effect of metallicity on TRGB magnitudes is illustrated by the black lines in the figures.
At fixed ${\rm [\alpha/Fe]}$, increasing metallicity leads to a fainter $I$ and $F814W$ TRGB luminosity in both sets of isochrones.
This behavior arises because higher metallicity enhances the efficiency of the CNO-cycle in the hydrogen burning shell, resulting in slightly earlier triple-$\alpha$ ignition at a lower helium core mass.
This metallicity dependence is also reflected in theoretical and empirical TRGB calibrations, which are expressed in various polynomial forms as functions of ${\rm [Fe/H]}$ and $(V-I)$ \citep[see Table~7 and Figure~13 of][]{2017ApJ...835...28J}.

As shown by the solid lines, $Y^2$-isochrones with $\alpha$-element enhanced models (${\rm [\alpha/Fe]} = 0.3$) generally yield similar or slightly fainter TRGB magnitudes than scaled-solar models (${\rm [\alpha/Fe]} = 0.0$) in the intermediate and metal-rich regimes. 
At fixed ${\rm [Fe/H]}$, the fainter TRGB luminosity in $\alpha$-enhanced populations, especially for ${\rm [Fe/H]} > -1.5$, reflects a reduced helium core mass regulated by the increased total metallicity. 
Enhanced $\alpha$-element abundances increase conductive opacity in the electron-degenerate helium core, reducing cooling efficiency and triggering helium ignition at a lower core mass \citep{2013A&A...558A..12V, 2013A&A...549A..50V}. 
The associated increase in mean molecular weight also modifies the stellar interior structure and shortens the evolutionary timescale, similar to the effects seen in helium-rich populations \citep{2011ApJ...740L..45C, 2017ApJ...842...91C}, contributing to earlier core ignition and fainter TRGB luminosity. 
In addition, $\alpha$-enhancement boosts hydrogen shell burning efficiency through the CNO-cycle, mainly due to increased oxygen abundance, which allows the shell to maintain energy balance at lower temperatures and further reduces TRGB brightness at fixed ${\rm [Fe/H]}$.
This trend reverses at lower metallicities of ${\rm [Fe/H]}\sim -2.5$ for $Y^2$-isochrones, whereas the BaSTI isochrones show brighter TRGB magnitudes for $\alpha$-enhanced models (${\rm [\alpha/Fe]}=0.4$) below ${\rm [Fe/H]} < -1.5$. 
This reversal is attributed to changes in atmospheric opacity and bolometric corrections specific to BaSTI models with ${\rm [\alpha/Fe]} = 0.4$ \citep{2006ApJ...642..797P}. 
Furthermore, in extremely metal-poor cases, the reduced total metallicity decreases opacity in the hydrogen shell, offsetting the effects of a smaller helium core. 
The enhanced CNO-cycle activity in these conditions further mitigates the dimming seen at higher metallicities, producing slightly brighter TRGB magnitudes.
Our comparisons, showing increased TRGB luminosity with higher ${\rm [\alpha/Fe]}$ in metal-poor models, align with \citet{2023AJ....166..224M}. 
Note that, as shown in Figure~\ref{f1}, the impact of $\alpha$-enhancement on TRGB magnitudes in $M_I$ and $M_{F814W}$ is small but measurable ($\le 0.03$~mag), particularly, in the metal-poor regime where precise TRGB measurements are critical for distance determinations, and these effects are closely linked to the well-known metallicity systematics in TRGB luminosity.

Figure~\ref{f2} illustrates the effect of stellar population age on the TRGB magnitude. 
As shown, in the metal-poor regime of ${\rm [Fe/H]} < -0.8$, the TRGB becomes slightly brighter for younger populations of 9~Gyr in the BaSTI models, and remains nearly the same or becomes only slightly brighter in the $Y^2$ models, at fixed metallicity.
In principle, the helium core mass at the onset of the helium flash increases with stellar age, indicating that younger stars have smaller core masses at ignition and thus lower TRGB luminosities. 
However, for relatively old stellar populations (ages greater than 4 Gyr), this age dependence becomes weak, making the TRGB luminosity nearly independent of age \citep[see Figure 7 of][]{2012A&A...547A...5V}. 
Consistent with this, our results show that the TRGB magnitude shift due to age is relatively minor in both $M_I$ and $M_{F814W}$ compared to the variations induced by differences in metallicity or ${\rm [\alpha/Fe]}$. 

\begin{deluxetable}{lc}
\tabletypesize{\scriptsize}
\tablewidth{0pt}
\tablecaption{\label{tab1} Input parameters adopted for the synthetic composite CMDs}
\tablehead{\colhead{Parameters} &\colhead{Adopted values}}
\startdata
Salpeter initial mass function, $s$ &$2.35$\\
Limit magnitude in $M_I$ and $M_{\rm F814W}$ & $7.0$~mag\\
Coefficients for error simulation & $a=0.01$ and $b=0.12$\\
Metallicity coverage in ${\rm [Fe/H]}$ & $-2.5$ to $0.5$ \\
Gaussian mean metallicity, $\left< {\rm [Fe/H]}\right>$ & $-1.5$, $-1.0$, and $-0.5$\\
$\alpha$-element enhancement, $[\alpha/{\rm Fe}]$& $0.0$ and $0.3$\\
Age, $t$ (Gyr) & 9.0 and 12.0\\
Initial helium abundance, $Y_{\rm ini}$ & $0.23$ and $0.33$\\
\enddata
\end{deluxetable}

Figures~\ref{f3} and \ref{f4} demonstrate the effect of helium enhancement on the TRGB magnitude in the $(V-I)_0$ vs. $M_I$ and $({{F606W} - F814W})_0$ vs. $M_{F814W}$ CMDs, respectively.
Helium-enhanced populations exhibit fainter TRGB magnitudes than normal-helium populations with $Y_{\rm ini} = 0.23$ in the metal-poor regime (${\rm [Fe/H]}<-1.6$). 
This behavior arises from more rapid stellar evolution driven by the increased mean molecular weight in the stellar interior \citep{2011ApJ...740L..45C, 2013ApJS..204....3C, 2017ApJ...842...91C, 2020ApJS..250...33C}. 
The higher mean molecular weight accelerates hydrogen shell burning, leading to a faster rise in core temperature relative to mass accumulation from the shell. 
As shown in the figures, this effect becomes pronounced for populations with an initial helium abundance of $Y_{\rm ini}=0.43$ (representing the most extreme enhancement, but confined to a minor subpopulation, as inferred for $\omega$~Cen and NGC~2808\footnote{The absolute TRGB magnitudes of $\omega$~Cen ($M_I = -4.028$) and NGC~2808 ($M_I = -4.239$), as derived from \emph{Gaia}, do not fully reflect the well-established metallicity-dependent systematics \citep{2021MNRAS.505.5957B}. Moreover, recent studies of Milky Way globular cluster distances based on \emph{Gaia} parallaxes, in combination with TRGB luminosities derived from \emph{Gaia} and \emph{HST} data, reveal a significant discrepancy in the absolute TRGB calibration for $\omega$~Cen \citep[i.e.,][]{2023ApJ...950...83L, 2025ApJ...980..218S}, suggesting that further investigation is still required.}), for which the TRGB in both $M_I$ and $M_{F814W}$ is 0.1–0.2 mag fainter than in normal-helium populations ($Y_{\rm ini}=0.23$) at the same metallicity in the metal-poor regime, assuming the system is entirely composed of $Y_{\rm ini}=0.43$ populations.

\section{Synthetic Composite Color Magnitude Diagrams and TRGB detection}
\label{s3}

Systematic variations in TRGB magnitude can be explored using synthetic CMDs, which enable controlled tests of how stellar parameters and observational uncertainties affect the TRGB luminosity.
We adopt the $Y^2$-isochrones for this analysis because they provide the full range of stellar population parameters required for this study, including metallicity (${\rm [Fe/H]}$), $\alpha$-element enhancements (${\rm [\alpha/Fe]}$), age (Gyr), and initial helium abundance ($Y_{\rm ini}$). 
Figure~\ref{f5} compares the TRGB magnitude trends with metallicity and $\alpha$-element enhancements from the $Y^2$, BaSTI, and MIST \citep {2016ApJ...823..102C} isochrones. 
Although the predicted absolute TRGB magnitudes differ somewhat among the models, particularly at low metallicity where the $Y^2$ models are relatively fainter and the MIST models brighter, the overall behavior remains very similar. We emphasize that the purpose of this study is not to establish an absolute calibration of $M_{I}^{\rm TRGB}$, as in \citet{2017ApJ...835...28J}, but rather to quantify the relative differences in TRGB magnitude arising from variations in stellar population parameters.
In particular, for the metallicity difference between ${\rm [Fe/H]}=-1.5$ and $-1.0$, the predicted TRGB magnitude differences in the $Y^2$ models are 0.058 and 0.119 mag for the $\alpha$-enhanced and scaled-solar cases, respectively, compared with 0.027 and 0.101 mag in BaSTI and 0.050 and 0.026 mag in MIST for the corresponding cases. 
These comparisons show that the differential trends are similar across the different stellar evolution models, and that the $Y^2$-based results provide a conservative reference framework for studying population-dependent variations in the TRGB magnitude.

\begin{figure*}
\centering
\includegraphics[angle=-90,scale=0.66]{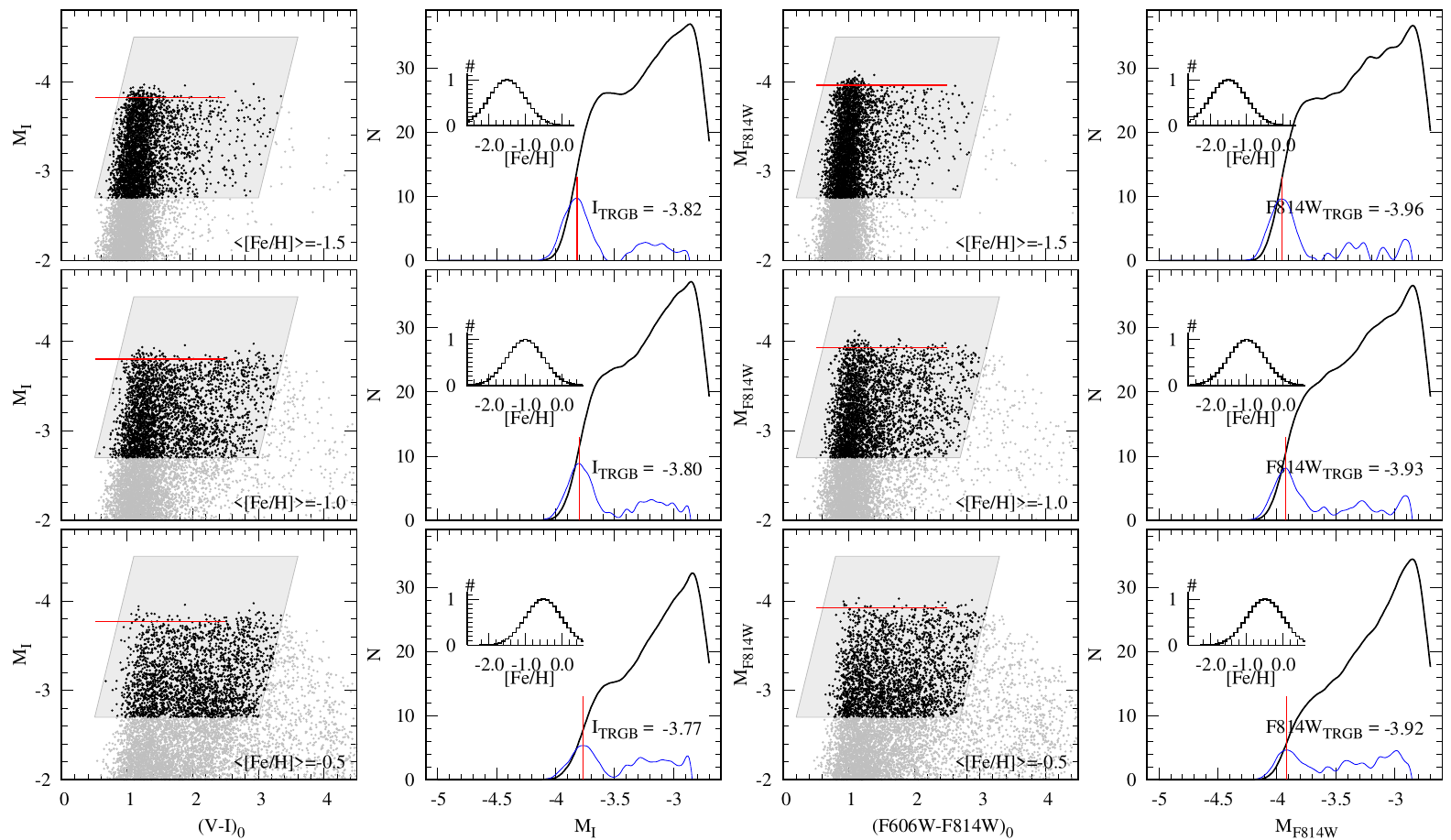}
\caption{
Same as Figure~\ref{f6}, but for the age of 9~Gyr. 
The TRGB magnitudes of the younger isochrones predict slightly brighter $M_I$ and $M_{F814W}$ magnitudes.
}
\label{f8}
\end{figure*}

\begin{figure*}
\centering
\includegraphics[angle=-90,scale=0.66]{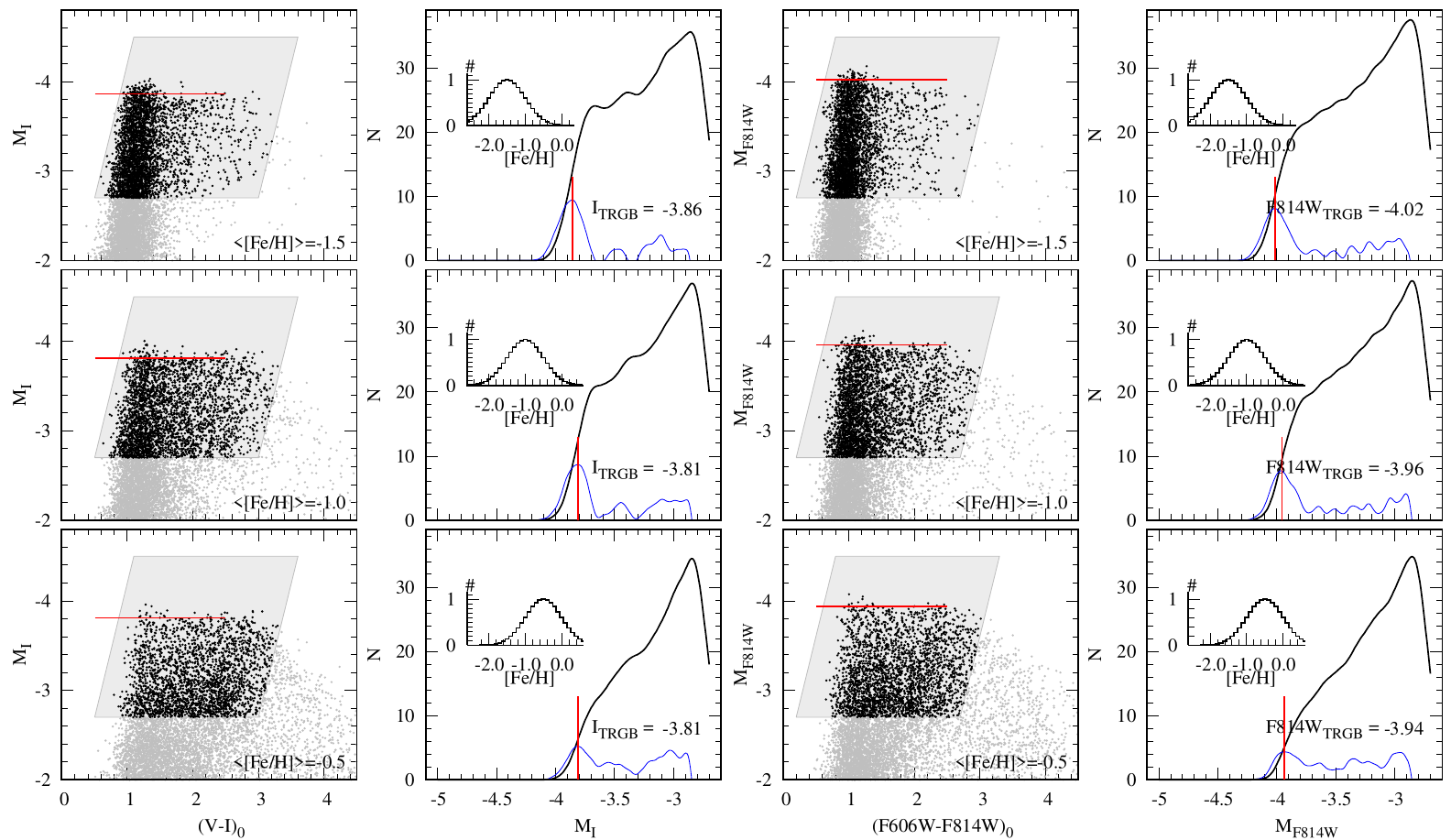}
\caption{
Same as Figure~\ref{f6}, but for the increased initial helium abundance of $Y_{\rm ini}=0.33$. 
The differences in $M_{I}^{\rm TRGB}$ and $M_{F814W}^{\rm TRGB}$ magnitudes are almost negligible compared to those of the normal-helium ($Y_{\rm ini} = 0.23$) populations. 
}
\label{f9}
\end{figure*}

\begin{figure*}
\centering
\includegraphics[angle=-90,scale=0.66]{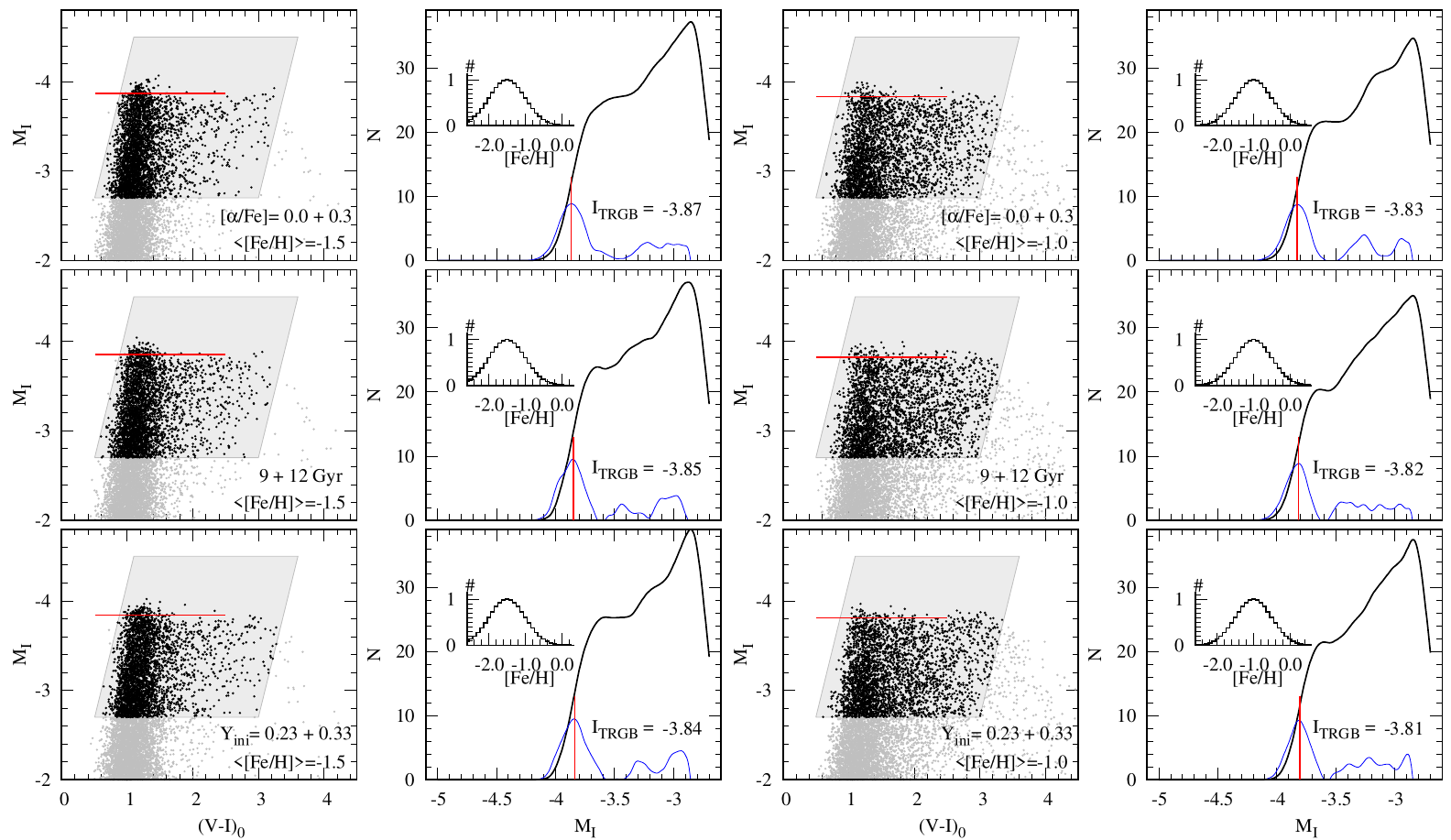}
\caption{
$M_{I}^{\rm TRGB}$ for mixtures of different stellar populations.
From top to bottom panels, the combinations include ${\rm [\alpha/Fe]} = 0.0$ and 0.3, ages of 9 and 12 Gyr, and stellar populations with $Y_{\rm ini} = 0.23$ and $0.33$. 
The left two columns correspond to $\left< {\rm [Fe/H]} \right> = -1.5$, while the right two columns correspond to $\left< {\rm [Fe/H]} \right> = -1.0$. 
The general trends for stellar population parameters identified in previous simulations remain consistent in these mixed-population cases.
}
\label{f10}
\end{figure*}

Based on the $Y^2$-isochrones, we construct synthetic composite CMDs for $(V-I)_0$ versus $M_I$ and $({{F606W} - F814W})_0$ versus $M_{F814W}$ using specified stellar population parameters \citep[see also][]{2013ApJS..204....3C, 2017ApJ...842...91C}. 
To simulate realistic TRGB observations, we assume that galaxy halos or outer regions, which are typically used for TRGB detection, are composed of old composite stellar populations with a range of metallicities. 
Accordingly, we introduce a metallicity spread by adopting a Gaussian metallicity distribution function (MDF).
For our synthetic composite CMDs, we adopt a conservative metallicity dispersion of $\sigma_{\rm [Fe/H]} = 0.5$, inferred from the M31 stellar halo \citep[][]{2010ApJ...708.1168T}, and consider three cases with mean metallicities of $\left< {\rm [Fe/H]} \right> = -1.5$, $-1.0$, and $-0.5$. 
On top of these MDFs, we compare the effects of varying ${\rm [\alpha/Fe]}$, stellar population age, and initial helium abundance $Y_{\rm ini}$, as shown in Figures~\ref{f6} to \ref{f10}.
Table~\ref{tab1} provides a summary of the input parameters adopted for the synthetic composite CMDs in this paper.

We first construct synthetic CMDs for the aforementioned two colors with metallicity steps of 0.1~dex, ranging from ${\rm [Fe/H]} = -2.5$ to $0.5$. 
For realistic simulations, we include photometric observational errors following $\sigma_{{\rm mag},i} = a \times 10^{b \times ({\rm mag},i - {\rm mag_{lim}})}$, where ${\rm mag},i$ denotes the simulated magnitude of each star and ${\rm mag_{lim}}$ is the limit magnitude of the observation. 
Using this $1\sigma$ Gaussian error at each magnitude, we assign photometric uncertainties to individual stars. 
In our simulation, we do not include asymptotic giant branch (AGB) stars, which can appear above the TRGB and potentially bias edge detection. 
For the old, halo-like stellar populations considered here, however, the RGB-to-AGB transition does not significantly affect TRGB detection \citep{2023AJ....166....2M}.
Each CMD at a given metallicity contains at least $\sim 1700$ stars for metal-poor populations and $\sim 5000$ stars for metal-rich populations brighter than $M_I \sim 2$. 
Based on these CMDs, we randomly select stars across metallicities to reproduce the assumed Gaussian MDF, resulting in approximately 16,000 stars per one synthetic composite CMD. 

To derive accurate TRGB magnitudes for each parameter set, we repeat the random composite CMD selection 100 times and apply the Sobel edge-detection algorithm to each composite CMD to obtain the mean TRGB magnitude along with the associated statistical TRGB magnitude scatter.
The Sobel edge-detection method is commonly used to identify the sharp discontinuity in the luminosity function of RGB stars \citep[e.g.,][]{1996ApJ...461..713S, 1997ApJ...480..589S}.
Here the Sobel filter is applied to the one-dimensional luminosity function derived from the CMD. 
The choice of smoothing scale and kernel controls the trade-off between noise suppression and edge sharpness.
We first smooth the luminosity functions of the simulated CMDs using 0.01~magnitude bin width, then apply the Sobel kernel $[2, 1, 0, -1, -2]$ to enhance the robustness of edge detection.

Figure~\ref{f6} illustrates the effect of increasing the mean metallicity of Gaussian MDF on the TRGB magnitudes. 
The CMDs shown represent a single example out of the 100 realizations of composite CMDs.
Throughout this paper, we do not apply color-based corrections to the TRGB magnitude in order to isolate the impact of the stellar parameters of interest.
As discussed in Section~\ref{s2}, the $M_I^{\rm TRGB}$ magnitude becomes slightly fainter with increasing metallicity, and this general trend in the isochrones is well reproduced in the synthetic composite CMDs. 
An increase in the mean $\left< {\rm [Fe/H]} \right>$ from $-1.5$ to $-1.0$ results in fainter$M_I^{\rm TRGB}$ luminosity by ${0.046 \pm 0.018}$~mag in our simulations.
Similarly, the $M_{F814W}^{\rm TRGB}$ magnitude at the same condition also becomes fainter with increasing metallicity by ${0.093 \pm 0.026}$~mag, consistent with the trend seen in the isochrones.
However, for the most metal-rich case with $\left< {\rm [Fe/H]} \right> = -0.5$, both $M_{I}^{\rm TRGB}$ and $M_{F814W}^{\rm TRGB}$ exhibit large uncertainties in TRGB detection. 
This is primarily due to the contribution of metal-rich stars in the Gaussian MDF, which weakens the sharpness of the $M_{I}^{\rm TRGB}$ edge in the Sobel response.
For $M_{F814W}^{\rm TRGB}$, the contribution of metal-rich stars in the MDF also contaminates the luminosity function by introducing much fainter stars, which affect the Sobel edge-detection response.
Our results are consistent with previously reported empirical TRGB calibrations that describe the metallicity dependence of the TRGB magnitude \citep[e.g.,][]{2004A&A...424..199B, 2009ApJ...690..389M, 2017ApJ...835...28J}.
The results for our simulations are summarized in Tables~\ref{t1} and \ref{t2}, which present the edge detection from 100 simulated realizations.

To compare with the ${\rm [\alpha/Fe]} = 0.3$ case, we present synthetic composite CMDs with scaled-solar $\alpha$-element abundances (${\rm [\alpha/Fe]} = 0.0$) in Figure~\ref{f7}. 
As discussed in Section~\ref{s2}, the effect of $\alpha$-element enhancement varies with ${\rm [Fe/H]}$. 
Although the TRGB of $Y^2$-isochrones with ${\rm [\alpha/Fe]} = 0.3$ shows slightly brighter $M_I$ magnitudes in the metal-poor regime, our simulations based on a Gaussian MDF increase the contribution from metal-rich populations in the synthetic composite CMDs. 
As a result, our synthetic composite CMDs show brighter TRGB magnitudes for populations with ${\rm [\alpha/Fe]} = 0.0$ compared to those with ${\rm [\alpha/Fe]} = 0.3$ under the same mean metallicity of the MDF.
At fixed ${\rm [Fe/H]}$, populations with ${\rm [\alpha/Fe]} = 0.0$ are more metal-poor in total metallicity, leading to brighter $M_I^{\rm TRGB}$ magnitudes than their $\alpha$-element enhanced counterparts. 
When the mean metallicities of the Gaussian MDF are $\left< {\rm [Fe/H]} \right> = -1.5$ and $-1.0$, the $M_I^{\rm TRGB}$ magnitude differences between ${\rm [\alpha/Fe]} = 0.0$ and 0.3 are $0.053 \pm 0.019$ and $0.047 \pm 0.012$~mag, respectively.
The $M_{F814W}^{\rm TRGB}$ varies by $ {0.031 \pm 0.023}$ and ${0.057 \pm 0.034}$ at $\left< {\rm [Fe/H]} \right> = -1.5$ and $-1.0$, respectively, for changes in ${\rm [\alpha/Fe]}$, as shown in Table~\ref{t2}. 
These variations are comparable to those produced by a 0.5~dex increase in $\left<{\rm [Fe/H]}\right>$, indicating that $\alpha$-element enhancement has a similarly strong effect on TRGB magnitudes through changes in total metallicity.

\begin{table*}
\centering
\caption{Mean (median) $M_{I}^{\rm TRGB}$ with statistical $1\sigma$ uncertainties estimated from Sobel edge detections based on 100 bootstrap resamplings.}
\begin{tabular}{ccccc}
\hline
{Metallicity}             & \multicolumn{4}{c}{Stellar parameters (${\rm [\alpha/Fe]}$, Gyr, $Y_{\rm ini}$)}     \\ 
\cline{2-5}
$\left < {\rm [Fe/H]} \right >$ & (0.3, 12.0, 0.23) & (0.0, 12.0, 0.23) & (0.3, 9.0, 0.23) & (0.3, 12.0, 0.33) \\ 
\hline
$-1.5$ & $-3.842\,(-3.840) \pm 0.016$ & $-3.895\,(-3.900) \pm 0.010$ & $-3.866 \, (-3.865) \pm 0.015$ & $-3.830\,(-3.830) \pm 0.009$ \\
$-1.0$ & $-3.796\,(-3.800) \pm 0.009$ & $-3.843\,(-3.840) \pm 0.008$ & $-3.834 \, (-3.830) \pm 0.010$ & $-3.801\,(-3.800) \pm 0.008$ \\
$-0.5$ & $-3.611\,(-3.770) \pm 0.312$ & $-3.821\,(-3.820) \pm 0.012$ & $-3.733\, (-3.810) \pm 0.237$ &  $-3.741\,(-3.790) \pm 0.173$ \\ 
\hline
\end{tabular}
\label{t1}
\end{table*}

\begin{table*}
\centering
\caption{Mean (median) $M_{F814W}^{\rm TRGB}$ with statistical $1\sigma$ uncertainties estimated from Sobel edge detections based on 100 bootstrap resamplings.}
\begin{tabular}{ccccc}
\hline
{Metallicity}             & \multicolumn{4}{c}{Stellar parameters (${\rm [\alpha/Fe]}$, Gyr, $Y_{\rm ini}$)}     \\ 
\cline{2-5}
$\left < {\rm [Fe/H]} \right >$ & (0.3, 12.0, 0.23) & (0.0, 12.0, 0.23) & (0.3, 9.0, 0.23) & (0.3, 12.0, 0.33) \\ 
\hline
$-1.5$ & ${-3.999\,(-4.000) \pm 0.013}$ & ${-4.030\,(-4.030) \pm 0.019}$ & ${-4.011\,(-4.010) \pm 0.008}$ & ${-3.963 \, (-3.960) \pm 0.010}$ \\
$-1.0$ & ${-3.906\,(-3.900) \pm 0.023}$ & ${-3.963\,(-3.970) \pm 0.025}$ & ${-3.963\,(-3.960) \pm 0.011}$ & ${-3.923 \, (-3.920) \pm 0.009}$ \\
$-0.5$ & ${-3.577\,(-3.850) \pm 0.363}$ & ${-3.683\,(-3.890) \pm 0.404}$ & ${-3.528\,(-3.895) \pm 0.481}$ & ${-3.827\, (-3.910) \pm 0.263}$ \\ 
\hline
\end{tabular}
\label{t2}
\end{table*}

\begin{table*}
\centering
\caption{Mean and median $M_{I}^{\rm TRGB}$ for mixed stellar populations, with statistical $1\sigma$ uncertainties estimated from Sobel edge detections using 100 bootstrap resamplings.}
\begin{tabular}{ccccc}
\hline
{Metallicity}             & \multicolumn{4}{c}{Standard model (${\rm [\alpha/Fe]}=0.3$, 12~Gyr, $Y_{\rm ini}=0.23$)}     \\ 
\cline{2-5}
$\left < {\rm [Fe/H]} \right >$ & $100\%$ Standard model & $50\% \, {\rm [\alpha/Fe]}=0.0$ & $50\% \,$ 9~Gyr & $50\% \, Y_{\rm ini}=0.33$ \\ 
\hline
$-1.5$ & $-3.842\,(-3.840) \pm 0.016$ & $-3.871\, (-3.870) \pm 0.019$ & $-3.860\, (-3.860) \pm 0.016$  & $-3.836\, (-3.830) \pm 0.016$ \\
$-1.0$ & $-3.796\,(-3.800) \pm 0.009$ & $-3.823\, (-3.820) \pm 0.017$ & $-3.818\, (-3.810) \pm 0.015$  & $-3.800\, (-3.800) \pm 0.012$ \\
$-0.5$ & $-3.611\,(-3.770) \pm 0.312$ & $-3.763\, (-3.800) \pm 0.160$ & $-3.596\, (-3.780) \pm 0.340$  & $-3.649\, (-3.780) \pm 0.296$ \\
\hline
\end{tabular}
\label{t3}
\end{table*}

We present the effect of age variation on TRGB magnitude in Figure~\ref{f8}.
Galaxy halos are among the oldest components, typically formed within the first few Gyr of a galaxy's history. 
Therefore, it is reasonable to assume that TRGB detections generally trace very old stellar populations, around 12 to 13 Gyr, with minimal age variation.
However, as described by \citet{1978ApJ...225..357S} and supported by \citet[][]{1994ApJ...423..248L}, the assembly history of galaxies may introduce an age structure in their halos, with potential age differences of several Gyr. 
Motivated by this, we test a 9~Gyr model in our synthetic composite CMDs for the comparison of 12~Gyr model.
In general, younger stars have smaller helium core masses, which would lead to slightly fainter TRGB luminosities at helium ignition. 
However, as shown in Figure~\ref{f8}, the 9~Gyr populations yield slightly brighter TRGB magnitudes compared to the 12~Gyr models. 
This behavior reflects the fact that, at fixed metallicity, younger stellar populations exhibit bluer colors than older populations, mimicking slightly more metal-poor star and producing a brighter TRGB magnitude.
As a result, slightly brighter TRGB magnitudes are observed in the younger models (see Figure~\ref{f2}). 
Nevertheless, as summarized in Tables~\ref{t1} and \ref{t2}, the $M_I^{\rm TRGB}$ magnitude differences for $\left< {\rm [Fe/H]} \right> = -1.5$ and $-1.0$ are $0.024 \pm 0.023$ and $0.038 \pm 0.013$~mag, respectively.
For $M_{F814W}^{\rm TRGB}$, the magnitude differences are comparable to those in $M_I^{\rm TRGB}$, yielding ${0.012 \pm 0.015}$ and ${0.057 \pm 0.025}$ under the same conditions. 
These differences are minor relative to the impact of metallicity or ${\rm [\alpha/Fe]}$ and lie within the typical scatter of TRGB detection, indicating that the age effect is comparatively less significant.

\begin{figure*}
\centering
\includegraphics[angle=-90,scale=0.66]{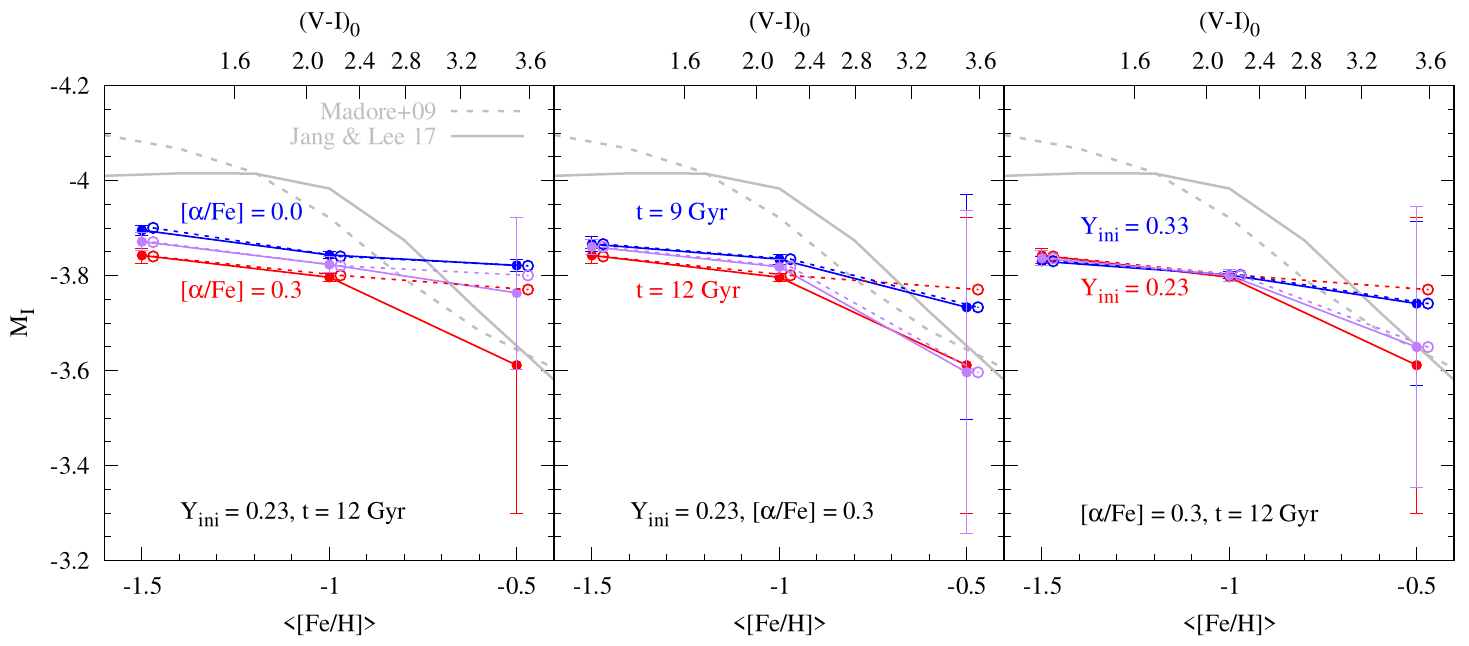}
\caption{
The effect of different stellar population parameters on the $M_{I}^{\rm TRGB}$.
From left to right, the panels show the influence of ${\rm [\alpha/Fe]}$, age, and $Y_{\rm ini}$, respectively. 
In each panel, red symbols represent the standard model sets, while blue and purple symbols correspond to comparison models for each parameter. 
Purple indicates a mixture of the two models with equal fractions. 
The simulations are based on synthetic composite CMDs generated at fixed stellar parameters, with $\pm 1\sigma$ scatter estimated from 100 random realizations using Sobel-edge detection. 
Solid circles denote mean TRGB magnitudes, and open circles indicate median values.
For comparison with empirical TRGB calibrations, the relations from \citet{2009ApJ...690..389M} and \citet{2017ApJ...835...28J} are shown as dashed and solid gray lines, respectively.
The upper x-axis gives the $(V-I)_0$ color corresponding to the ${\rm [Fe/H]}$.}
\label{f11}
\end{figure*}

The effect of helium enhanced stellar populations on the TRGB magnitude is presented in Figure~\ref{f9}.
Considering the typical helium enhancement predicted by \citet{2018MNRAS.481.5098M}, second-generation populations observed in Milky Way globular clusters show an average helium enhancement of $\Delta Y \le 0.05$, which corresponds to $Y_{\rm ini} = 0.28$ in the $Y^2$-isochrones. 
If such populations are mixed into galaxy halos, the resulting helium enhancement would lead to only minor variations compared to normal-helium populations with $Y_{\rm ini} = 0.23$. 
However, to examine the maximal impact of helium enhancement on the TRGB magnitude, we adopt $Y_{\rm ini} = 0.33$. 
This choice is motivated by some globular clusters that host significantly helium-enhanced subpopulations, such as $\omega$~Cen, NGC~2808, NGC~6388, and NGC~6441 \citep[e.g.,][]{2005ApJ...621L..57L, 2007ApJ...661L..53P, 2008ApJ...677.1080Y}.
Although the most extremely helium-enhanced stellar populations in these globular clusters reach $Y_{\rm ini} > 0.38$, their number fraction is less than $\sim10\%$, making $Y_{\rm ini} = 0.33$ a representative value for our TRGB analysis.
As discussed in Section~\ref{s2}, increased helium abundance reduces the core mass at the helium flash, leading to a lower TRGB luminosity and consequently a fainter absolute magnitude compared to normal-helium populations. 
However, as the helium abundance increases, the hydrogen mass fraction decreases, leading to a lower total metallicity at a fixed ${\rm [Fe/H]}$. 
This produces a counteracting effect on the helium core mass despite the higher helium abundance.
As a result, the helium-enhanced population with $Y_{\rm ini} = 0.33$ shows negligible variation in TRGB magnitudes, with differences in $M_I$ and $M_{F814W}$ falling within the observed scatter. 

With models incorporating various stellar parameters, we are now able to investigate the impact of mixed stellar populations on TRGB magnitudes. 
We tested three cases involving mixtures in ${\rm [\alpha/Fe]=0.0}$, stellar population age of $t=9$~Gyr, and initial helium abundance of $Y_{\rm ini}=0.33$. 
Figure~\ref{f10} presents simulations where two populations are combined in equal proportions with the base model of ${\rm [\alpha/Fe]}=0.3$, $t=12$~Gyr, and $Y_{\rm ini}=0.23$. 
The mixtures are synthetic composite CMDs for ${\rm [Fe/H]} = -1.5$ and $-1.0$. 
As summarized in Table~\ref{t3}, the TRGB magnitudes from the mixed populations closely match the averaged TRGB magnitudes from the individual components. 
Therefore, the resulting magnitude differences due to the mixing effect of different populations are small compared to those driven by changes in metallicity or ${\rm [\alpha/Fe]}$ alone.

Figure~\ref{f11} summarizes the TRGB detection results from our simulations.
For reference, we also include the empirical TRGB calibrations of \citet{2009ApJ...690..389M} and \citet{2017ApJ...835...28J}.
As noted above, our synthetic composite CMDs are not intended to reproduce the absolute TRGB calibration.
Rather, we focus on the relative dependence of the TRGB magnitude on metallicity.
A detailed discussion of the differences between theoretical isochrones and empirical TRGB calibrations is presented in Section~\ref{f4}.
Empirical TRGB calibrations predict $0.03 \lesssim \Delta M^{\rm TRGB}_I \lesssim 0.16$ for a metallicity change from ${\rm [Fe/H]}=-1.5$ to $-1.0$, and the TRGB magnitude variations shown in the figure are broadly consistent with these expectations for the adopted stellar parameters.
The effect of population mixing is most pronounced when scaled solar models are included, while variations in age or helium enhanced populations are largely negligible.
Except for the case of ${\rm [Fe/H]} = -0.5$, the influence of population mixing on the TRGB magnitude remains modest. 
When multiple parameters that affect the TRGB magnitude, such as metallicity, ${\rm [\alpha/Fe]}$, and age, act in the same direction, their combined effect may become significant.
Nevertheless, typical TRGB observations focus on relatively homogeneous stellar populations in the old, metal-poor regions of a galaxy's halo. 
Under these conditions, even with some variation in helium abundance or age within the halo \citep[e.g.,][]{2016MNRAS.456L...1C, 2019ApJ...883L..31C}, the overall impact of population differences on the TRGB magnitude is expected to be negligible.

\section{Summary and Discussion}
\label{s4}

We have used synthetic composite CMDs to examine how metallicity, $\alpha$-element enhancement, stellar age, and helium abundance variation affect TRGB magnitudes in the $I$ and $F814W$ bands. 
As expected, increasing metallicity leads to fainter TRGB magnitudes, both in $M_I$ and $M_{F814W}$, consistent with previous theoretical and observational studies.
Given typical metallicity uncertainties of 0.2–0.4~dex in the halo, our simulation is consistent with the systematic errors in TRGB distance measurements, corresponding to a 0.02–0.04~mag shift in $M_I^{\rm TRGB}$ \citep[e.g.,][]{2017ApJ...835...28J}.
At fixed ${\rm [Fe/H]} = -1.0$, an increase in ${\rm [\alpha/Fe]}$ from 0.0 to 0.3 produces a measurable $M_I^{\rm TRGB}$ magnitude difference of $0.047 \pm 0.012$ mag. This reflects the influence of $\alpha$-elements on total metallicity and helium core mass.
The effect of age is slightly weaker. 
A 3~Gyr decrease from 12~Gyr at ${\rm [Fe/H]} = -1.0$ results in a $0.038 \pm 0.013$~mag shift, representing a lower age bound for typical halo populations. 
If the age difference in the halo population is less than 1 Gyr, the resulting variation would remain smaller than those caused by metallicity or $\alpha$-element differences.
The impact of helium variation is smaller still, yielding only $0.005 \pm 0.012$ mag difference between $Y_{\rm ini} = 0.23$ and $0.33$, even though helium strongly affects broadband colors at shorter wavelengths.
Simulations of mixed populations show that TRGB magnitudes closely follow the average of their components. 

While our synthetic composite CMDs are used only to examine relative TRGB magnitude differences, it is important to note that a long-standing discrepancy persists between the observed TRGB magnitude and the absolute magnitudes predicted by theoretical models. 
Recent non-rotating calculations that incorporate atomic diffusion and updated conductive opacities have reduced this offset to $\sim 0.03$–$0.05$~mag, yet the predicted $I$-band TRGB luminosity still shows a residual discrepancy relative to observations \citep{2017A&A...606A..33S, 2021A&A...654A.149C}. 
Several explanations have been proposed, and stellar rotation provides a particularly natural solution. 
Rotation affects stellar evolution through two channels. 
First, moderate initial rotation, typical of old Population~II stars, enhances internal mixing during the main sequence and early RGB phases, increasing the helium core mass and altering envelope opacities \citep[e.g.,][]{2012A&A...537A.146E}.
Second, the centrifugal force partially counteracts gravity, delaying the helium core flash and allowing the core to grow to a larger mass before ignition. 
Both effects brighten the theoretical TRGB by $0.01$–$0.04$~mag \citep{2010A&A...522A..10C, 2012A&A...543A.108L, 2017A&A...597A..14G}.
Because the $Y^2$-isochrones used in our analysis do not include rotational physics, they cannot capture these absolute magnitude shifts. 
If these rotational effects were incorporated into the $Y^2$-evolutionary tracks, the predicted TRGB luminosity would increase and become consistent with empirically calibrated values.

Our results in this paper are particularly relevant to the ongoing debate over the $H_0$ determined from measurements in the nearby universe and at the epoch of the cosmic microwave background \citep[see][and references therein]{2021CQGra..38o3001D}.
Interestingly, this tension also exists in the nearby universe $H_0$ measurements based on TRGB- and Cepheid-based calibrations of Type Ia supernovae (SNe Ia). 
The Carnegie–Chicago Hubble Program \citep[CCHP;][]{2019ApJ...882...34F} uses the TRGB method, whereas the Supernovae and $H_0$ for the Equation of State program \citep[SH0ES;][]{2022ApJ...934L...7R} employs Cepheid variables. 
Both achieve $\sim$1–2\% internal precision, yet differences in their zero-point calibrations, caused by methodological and astrophysical factors, lead to a significant offset in $H_0$ \citep{2021ApJ...919...16F}.
SH0ES generally reports values around $73$ to $74\,{\rm km/s/Mpc}$, while CCHP finds $69$ to $70\,{\rm km/s/Mpc}$. 
The Comparative Analysis of TRGBs \citep[CATS;][]{2023ApJ...954L..31S} reports a similar $H_0$ values consistent with those from SH0ES, despite using TRGB luminosities. 
Even accounting for the $\sim0.03$~mag zero-point offset in absolute $M_I^{\rm TRGB}$ magnitude, the TRGB method alone is unlikely to reconcile these two $H_0$ scales.
However, once a unified calibration of the absolute TRGB magnitude is established, the demonstrated robustness of the TRGB method will strongly favor one interpretation of the $H_0$ tension over the other.
In particular, alignment with the CCHP calibration would largely reduce the $H_0$ discrepancy, whereas consistency with the SH0ES calibration would point to additional astrophysical or systematic effects, particularly those associated with the SN~Ia distance scale.

Studies of the SN~Ia host age bias \citep[e.g.,][]{2020ApJ...903...22L, 2022MNRAS.517.2697L, 2023ApJ...959...94C, 2025MNRAS.538.3340C, 2025MNRAS.544..975S} show that the SN~Ia distance scale depends on host age, whereas the TRGB magnitude, when measured in old, halo-dominated stellar populations, remains highly stable even when variations in stellar population properties, particularly age, are considered. 
If a measurable age difference exists between the nearby calibration and Hubble flow samples used for $H_0$ determination, it would not affect TRGB-based distances but could produce a detectable difference in the absolute magnitudes of SNe~Ia.
Interestingly, a population mismatch exists between the Cepheid calibrator sample of SNe~Ia and the Hubble flow sample. 
Cepheid calibrators are dominated by young stellar populations, whereas the Hubble flow sample includes about 20–30\% early-type galaxies that typically host older stellar populations.
If this morphological difference is confirmed to reflect an age difference through direct age dating, the high $H_0$ value from the SH0ES calibration would decrease, and the tension would be partially alleviated. 
Accurate age dating of both calibrators using a well-calibrated TRGB method, which is relatively less sensitive to age difference, could ultimately help resolve the current $H_0$ tension between the nearby and high-redshift universe (Chung et al. 2026 in prep.).

The demonstrated robustness of the TRGB method highlights the importance of refining the cosmic distance ladder through multiple independent and complementary approaches. 
Extending such calibrations to other distance indicators, such as surface brightness fluctuations that are sensitive to the helium abundance of stellar populations \citep[e.g.,][]{2020ApJS..250...33C, 2024ApJ...973...83A}, provides a pathway toward a unified and accurate distance scale across multiple standard candles, including SNe~Ia. 
Together, these efforts open new opportunities for precision cosmology and may help address ongoing tensions, including the long-standing $H_0$ discrepancy. 
In this context, robust TRGB measurements with the James Webb Space Telescope, combined with future megamaser distances determinations in the Hubble flow sample using the Event Horizon Telescope, hold significant promise for advancing our understanding of several key challenges in modern cosmology.

\begin{acknowledgments}
We thank the referee for a number of helpful comments and suggestions.
C.C., Y.-W.L., and S.-J.Y. acknowledge support from the National
Research Foundation (NRF) of Korea to the Center for Galaxy Evolution Research (RS-2022-R070872, RS-2022-NR070525).
S.-J.Y. acknowledges support from the Mid-career Researcher Program (RS-2024-00344283) through Korea's NRF funded by the Ministry of Science and ICT.
Y. -C.K. was supported by Basic Science Research Program through the NRF of Korea funded by the Ministry of Education, Science and Technology (NRF-2017R1D1A1B05028009).
S.-I.H. acknowledge support provided by the NRF of Korea grant funded by the Ministry of Science and ICT (RS-2021-NR058093)
D.L. acknowledges support from Basic Science Research Program through the NRF of Korea funded by the Ministry of Education (RS-2025-25419201).
Y.-L.K. was supported by the Lee Wonchul Fellowship, funded through the BK21 Fostering Outstanding Universities for Research (FOUR) Program (grant No. 4120200513819).
S.J. acknowledges support from the Postdoctoral Researcher Growth Program (RS-2025-25419519) funded by the NRF, under the Ministry of Education.
S.H. was supported by the Global-LAMP Program of the NRF grant funded by the Ministry of Education (RS-2023-00301976)
M.G.L. was supported by the NRF grant funded by the Korea government (MSIT) (RS-2024-00340832).
\end{acknowledgments}

\bibliography{trgb_ref}{}
\bibliographystyle{aasjournalv7}



\end{document}